# Foucault imaging and small-angle electron diffraction in controlled external magnetic fields

Hiroshi Nakajima[1], Atsuhiro Kotani[1], Ken Harada[1,2], Yui Ishii[1] and Shigeo Mori[1,*]

[1] Department of Materials Science, Osaka Prefecture University, Sakai, Osaka 599-8531, Japan
[2] RIKEN Center for Emergent Matter Science, Hatoyama, Saitama 350-0395, Japan
[*]To whom correspondence should be addressed. E-mail: mori@mtr.osakafu-u.ac.jp

We report a method for acquiring Foucault images and small-angle electron diffraction patterns in external magnetic fields using a conventional transmission electron microscope without any modification. In the electron optical system that we have constructed, external magnetic fields parallel to the optical axis can be controlled using the objective lens pole piece under weak excitation conditions in the Foucault mode and the diffraction mode. We observe two ferromagnetic perovskite-type manganese oxides, $La_{0.7}Sr_{0.3}MnO_3$ and $Nd_{0.5}Sr_{0.5}MnO_3$, in order to visualize magnetic domains and their magnetic responses to external magnetic fields. In rhombohedral-structured $La_{0.7}Sr_{0.3}MnO_3$, pinning of magnetic domain walls at crystallographic twin boundaries was found to have a strong influence on the generation of new magnetic domains in external applied magnetic fields.

**Introduction**

Magnetic materials contain various types of magnetic textures and magnetic domain structures that have certain influences on physical properties. For example, in manganese oxides with the perovskite structure, electronic phase separation of a ferromagnetic metal state and a charge-ordered insulating state is revealed to be a source of colossal magnetoresistance [1]. Recently, noncollinear spin structures such as spiral magnetism and magnetic skyrmions have been found to give rise to ferroelectricity and the topological Hall effect [2, 3]. However, applied magnetic fields are necessary to induce such magnetic spin structures in helical magnets. Thus, to uncover the mechanisms behind the physical properties of such materials, it is important to investigate variations in microscopic magnetic domain structures as functions of magnetic and electric fields, as well as temperature. Several imaging methods for visualizing magnetic domain structures in a real space are available, such as spin-polarized scanning tunneling microscopy [4], photoelectron emission microscopy [5, 6], spin-polarized scanning electron microscopy [7] and Lorentz microscopy. Among them, Lorentz microscopy has certain advantages for observing magnetic domains and their dynamic behavior in applied magnetic fields and different temperatures at nanoscale spatial resolution.

Furthermore, information about crystallographic symmetry and domain structures can be obtained using dark-field imaging and electron diffraction methods [8].

Lorentz microscopy uses two main imaging modes: Fresnel (out-of-focus) and Foucault (in-focus) modes [9]. The former can highlight domain walls in defocused conditions and is often used to obtain certain information about magnetic domain structures. The latter visualizes domains themselves by selecting spots deflected by the magnetization of a specimen. In order to obtain quantitative analyses of magnetic domain structures, such as the magnitude and orientation of magnetization in each domain, one of the most effective methods is to acquire small-angle electron diffraction (SmAED) patterns with a long camera length. The SmAED technique has been utilized in the past to observe striated structures [10], metallic films [11], and superconducting vortex lattices [12]. Recently, the SmAED method has been applied to magnetic domain observations in permalloy and chiral helimagnets such as $Cr_{1/3}NbS_2$ [13-15]. One of the most significant advances in SmAED that has been achieved is the simultaneous use of SmAED with Foucault imaging in a conventional transmission electron microscope (TEM) by converging the electron beam on the selected-area aperture plane [16, 17]. In the electron optical systems, an objective lens can be used to apply magnetic fields to a specimen. However, these methods have not previously included external magnetic fields for observations of magnetic materials, despite the importance of observing magnetic domain behavior in response to applied fields.

In this paper, we demonstrate a method of Foucault imaging and SmAED under external magnetic fields using an electron optical system. We apply the method to elucidate the static and dynamic behaviors of magnetic domains in the ferromagnetic phases of $La_{0.7}Sr_{0.3}MnO_3$ (LSMO) [18] and $Nd_{0.5}Sr_{0.5}MnO_3$ (NSMO) [19, 20] in applied magnetic fields. We observe distinct in-field behavior of magnetic domains when magnetic domain walls are pinned by crystallographic boundaries.

**Optical system**

We have constructed an electron optical system with an excited objective lens based on the previously reported electron optical system [17]. Figures 1(a) and 1(b) display schematic illustrations of the electron optical systems constructed for SmAED and Foucault imaging methods, respectively. These electron optical systems are controlled as follows. Condenser lenses are strongly excited in order to produce a divergence angle of less than $1 \times 10^{-6}$ rad in SmAED mode. By tuning the current value of an objective mini-lens, the lower crossover position can be adjusted on the selected-area aperture plane. When the objective lens is excited to apply external magnetic fields, the current of the objective mini-lens is decreased so as not to shift the crossover on the selected-area aperture plane. Note that the objective lens pole piece is used to apply external magnetic fields parallel to the optical axis. The objective aperture can be used to select a specimen area for obtaining SmAED patterns. The SmAED pattern is projected on a fluorescent screen by adjusting the current values of the image-forming lens. Camera lengths and magnifications depend on the current values of the image-forming lens. Switching from the SmAED to Foucault imaging mode is achieved by decreasing the current value of the intermediate lens 1 after selecting a magnetically deflected spot with the selected aperture.

**Methods**

**Sample preparation**

Single crystals of LSMO and NSMO were grown by a floating-zone method. Thin specimens of approximately 100 nm in thickness were prepared for TEM observation by Ar-ion milling. The thin specimens of LSMO and NSMO were mounted on double-tilt holders and tilted by a few degrees to reduce the contrast of bend contours and to secure wide fields of view for the observation. Observations were performed at room temperature and approximately 200 K for LSMO and NSMO, respectively.

**Construction of optical system**

In this work, Foucault imaging and SmAED were performed with a commercial 200 kV TEM (JEM-2010) equipped with a thermal-emission-type $LaB_6$ filament. The TEM comprises one objective mini-lens, three intermediate lenses, and one projection lens. The smallest selected-area aperture was 5 μm in diameter, which was used to select magnetically deflected spots, corresponding to 50 μrad in the reciprocal space in the optics set-up mentioned above.

The electron optical system was evaluated using a carbon grating with a lattice spacing of 500 nm in the same manner as described in Ref. [17, 21]. Fig. 2(a) shows camera lengths as a function of current values in the intermediate lenses 2 ($I_2$), 3 ($I_3$) and the objective lens ($I_{OL}$). Note that the intermediate lenses 2 and 3 are located below the intermediate lens 1. For convenience, the camera lengths used with the objective lens turned off are depicted in Fig. 2(a) based on the experimental results in Ref. [17]. The $I_3$ dependence of the camera length under external magnetic fields has a similar tendency as that of the camera length used without exciting the objective lens, although the camera lengths become much longer when the objective lens is excited. Note that $I_{OL} = 0.18$ and 0.36 A correspond to external magnetic fields with magnetic flux densities of 100 and 190 mT, respectively. Fig. 2(b) shows the $I_{OL}$ dependence of the current in the objective mini-lens ($I_{OM}$). When $I_{OL}$ is increased, $I_{OM}$ is decreased in order to maintain the crossover position on the selected-area aperture plane. Similarly, $I_{OM}$ is decreased when the condenser lens current ($I_{C3}$) is increased. SmAED patterns and Foucault images can be obtained at fields of up to 190 mT in the TEM. As shown in Fig. 2(b), it is possible to control the current in the objective lens (and the external magnetic fields) and the condenser lens (the specimen area irradiated by the electron beam) by adjusting the current of the objective mini-lens.

**Results and discussion**

**Observation of magnetic domains in LSMO**

In order to demonstrate the performance of the present electron optical system, we applied the method to the observation of magnetic domains in the crystallographic (111) plane of the ferromagnetic phase LSMO at room temperature. Fig. 3(a) shows an in-focus image acquired with the objective lens turned off. Fig. 3(b) displays an under-focused Fresnel image of 180° magnetic domains with magnetizations pointing to the right ($M_R$) and left ($M_L$). Judging from discontinuities in the bend contours, the 180° magnetic domains coincided with crystallographic twin boundaries (see arrowheads in Fig. 3(a)). The SmAED pattern of the specimen in Fig. 3(a) with a camera length of 150 m is shown in Fig. 3(c). The 000 direct spot was split by magnetic deflection due to the 180° magnetic domains with magnetic moments antiparallel between the two neighboring domains. A diffuse streak between the two split spots indicates the presence of Bloch-type magnetic domain walls in the 180° magnetic domains. The Foucault images obtained by selecting each of the two split spots (A and B) are shown in Figs. 3(d) and 3(e). The 180° magnetic domains can be seen clearly as regions of bright and dark contrast with the contrasts reversed in Figs. 3(d) and 3(e). The Lorentz deflection angle $\beta$ depends on both the in-plane magnetic flux density $B$ and the specimen thickness $t$. The angle is given by the equation $\beta = e\lambda Bt/h$, where $e$ is the electric charge, $\lambda$ is the wavelength of electron, and $h$ is the Planck constant [22]. The magnetic flux density produced by the specimen magnetization was evaluated to be approximately 0.5 T (2.6 $\mu_B$/Mn site) from the half angle ($\beta = 3.0 \times 10^{-5}$ rad) between spots A and B, assuming a specimen thickness of 100 nm.

We investigated changes in the magnetic domain structure of LSMO by applying an external magnetic field perpendicular to the plane of the thin specimen. Reconstruction of the magnetic domains was found to occur when a magnetic flux density of 50 mT was applied. Fig. 4(a) shows an under-focused Fresnel image in the identical region as in Fig. 3(b). When the magnetic field was applied parallel to the optical axis, new magnetic domains were generated, which are indicated by blue arrows. Comparing Figs. 3 and 4, we observed that the new domains appeared only in domains magnetized along $M_L$ and the domains with the opposite magnetization $M_R$ remain intact, indicating that the magnetic domain walls were strongly pinned by twin boundaries in the crystal. Moreover, it is suggested that magnetic domains with $M_L$ were unstable in external magnetic fields because the vertical component of magnetization $M_L$ may have become oriented antiparallel to the external magnetic field due to a small tilt in the specimen. These results suggest that the rhombohedral structure of LSMO has a weak magnetocrystalline anisotropy and is subject to changes in the magnetization direction in external applied fields. Fig. 4(b) is the SmAED pattern obtained in the circular region shown in Fig. 4(a). An extra spot appears in the SmAED pattern as indicated by C', in addition to the two split spots (A' and B') due to the 180° magnetic domains. The deviation angle of the new magnetization (blue arrows) from the original direction can be estimated from the SmAED pattern. The deviation angle was determined to be approximately $\theta = 22°$ by measuring the angle from the SmAED pattern. This angle is not related to the crystal orientation. Foucault images are shown in Figs. 4(c) and 4(d). In Fig. 4(c), two split spots (A' and C') were selected and two types of magnetic domains were visible. Conversely, the magnetic domains with $M_R$ appeared as regions of bright contrast in Fig. 4(d) when spot B' was selected. These Foucault images clearly show that the new domains curve around the crystallographic twin boundaries, which indicates that the generation of magnetic domains is likely affected by pinning at twin boundaries. Note that the directions of magnetization $M_R$ and $M_L$ remain unchanged although the magnetic domain walls appear curved.

### Observation of magnetic domains in NSMO

In addition, we investigated magnetic domains in the ferromagnetic phase NSMO at 200 K in fields of 0 and 170 mT. Figs. 5(a) and (b) are Fresnel and Foucault images with the objective lens turned off, illustrating 180° magnetic domain structures with in-plane magnetic moments. Note that the magnetic domain walls were not pinned by any crystal boundaries in the region, unlike in LSMO. A SmAED pattern with a camera length of 90 m is shown in the inset of Fig. 5(a). The 000 direct spot was split into two spots by magnetic deflection as in the case of LSMO. The deflection angle of NSMO at 200 K is $\beta = 3.5 \times 10^{-5}$ rad, which is nearly the same value as that of LSMO. According to magnetization measurements [18, 19], the magnetizations of both LSMO at 300 K and NSMO at 200 K are determined to be approximately 2.5 $\mu_B$/(Mn site) in a field of 0.5 T, which is consistent with the magnetizations obtained from the deflection angles in SmAED patterns. Fig. 5(b) shows a Foucault image obtained using the spot deflected by the magnetization labeled as $M_U$, in which the magnetic domains giving rise to the spot can be seen as regions of bright contrast. We observed magnetic domains in Fresnel and Foucault imaging modes at 170 mT. When the external magnetic field is applied perpendicular to the plane of the thin specimen of NSMO, the magnetic domains labeled as $M_D$ expanded and the others indicated by $M_U$ shrank, as shown in Figs. 5(c) and (d). This phenomenon is ascribed to the out-of-plane components of magnetization, and the out-of-plane direction of the magnetization $M_D$ is assumed to be parallel to the external magnetic field. Note that Fig. 5(d) was visualized by selecting the spot deflected by $M_D$, and this spot has a stronger intensity in the SmAED pattern. This behavior is different from that observed in LSMO, indicating that pinning of magnetic domain walls at crystallographic twin boundaries has an important influence on the behavior of LSMO. As demonstrated in the ferromagnetic phases LSMO and NSMO, we succeeded in obtaining Foucault images combined with their corresponding SmAED patterns in the electron optical system described here. In addition, the magnitude and direction of the magnetization in each magnetic domain can be determined and dynamic changes in magnetic domains due to external magnetic fields can be elucidated.

### Conclusions

In summary, we demonstrated the effectiveness of an electron optical system constructed for Foucault imaging and SmAED in external magnetic fields. To perform Foucault imaging, the objective mini-lens and condenser lens are adjusted such that the electron beam converges on the selected-area aperture plane. As a result, this method allows us to apply external magnetic fields with the objective lens and control the illumination and imaging systems independently. Magnetic domains of LSMO and NSMO in external magnetic fields were visualized by Foucault imaging using individual spots from SmAED patterns. An anomalous magnetic response was observed when magnetic domain walls are pinned at crystallographic twin boundaries in LSMO. This method will play a significant role in better understanding the physical properties of functional materials such as strongly correlated and magnetic materials.

ACKNOWLEDGMENTS

This work was partially supported by Grant-in-Aid (No. 16H03833 and No. 15K13306) from the Ministry of Education, Culture, Sports, Science and Technology (MEXT), Japan.

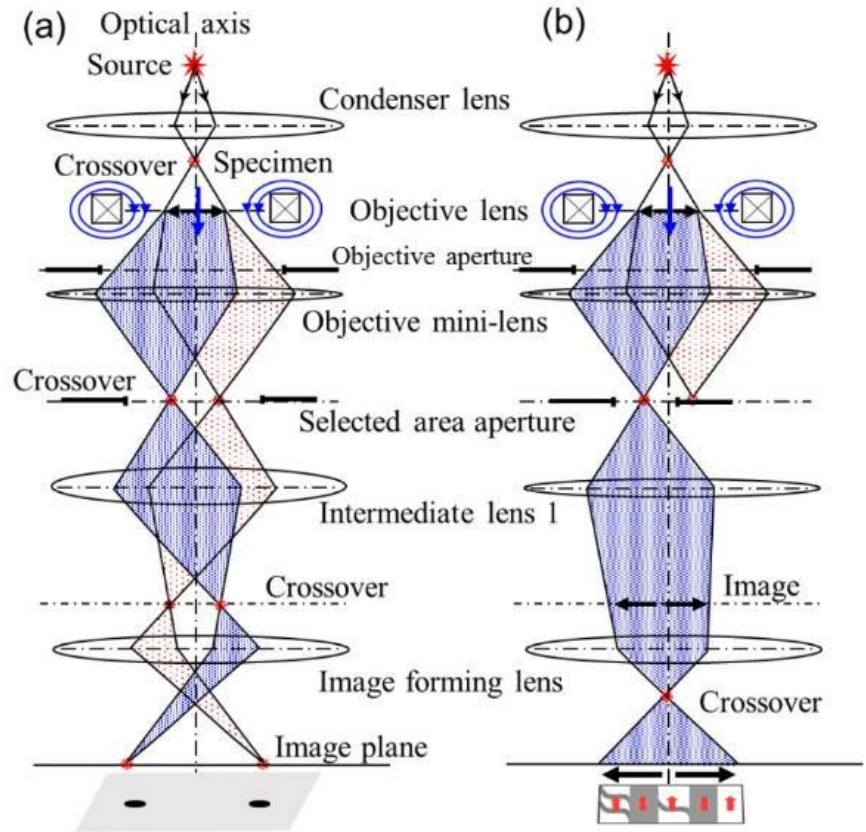

FIG. 1. Schematic illustrations of constructed electron optical systems with a weakly excited objective lens: (a) SmAED mode and (b) Foucault imaging mode.

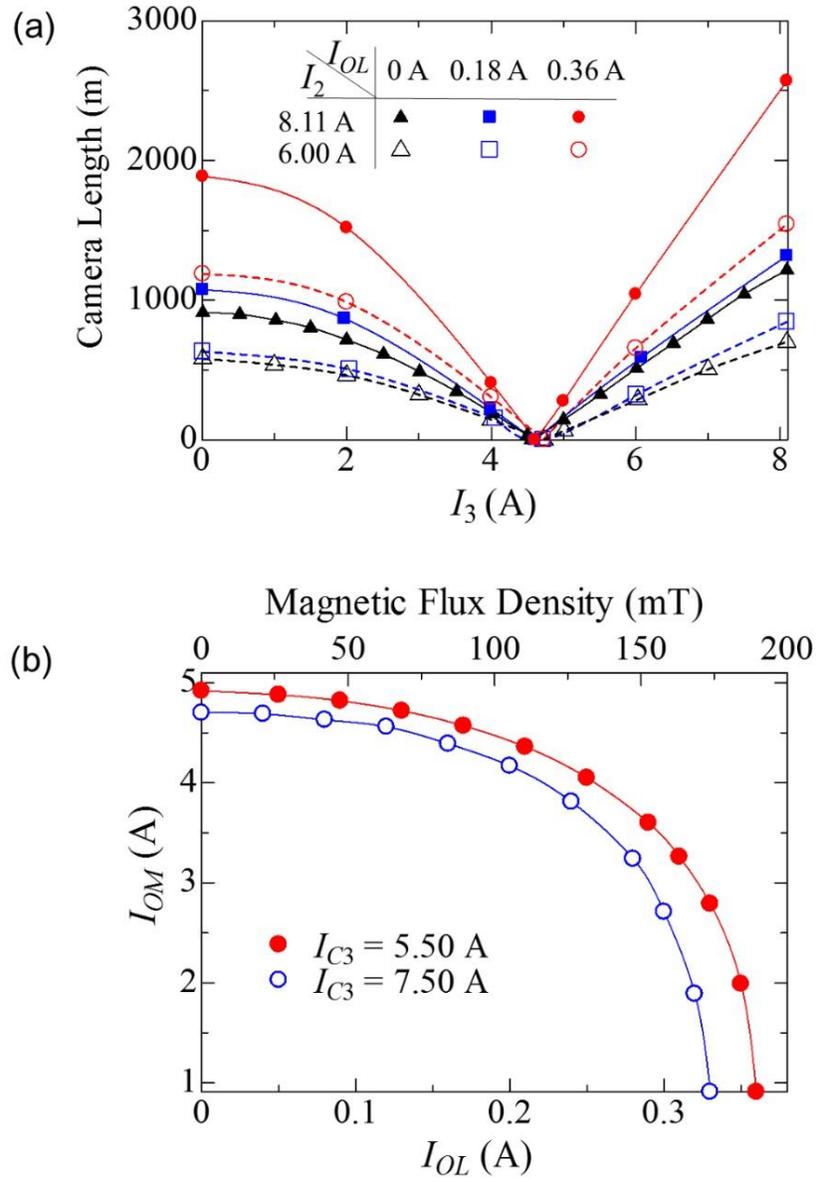

FIG. 2. (a) Current in intermediate lens 3 vs. camera lengths with and without the objective lens excited. $I_2$ and $I_3$ represent the current values in intermediate lenses 2 and 3, respectively. (b) The current in the objective lens ($I_{OL}$) vs. current in the objective mini-lens ($I_{OM}$). $I_{C3}$ represents the current value in condenser lens 3. Lines are provided to guide the eye.

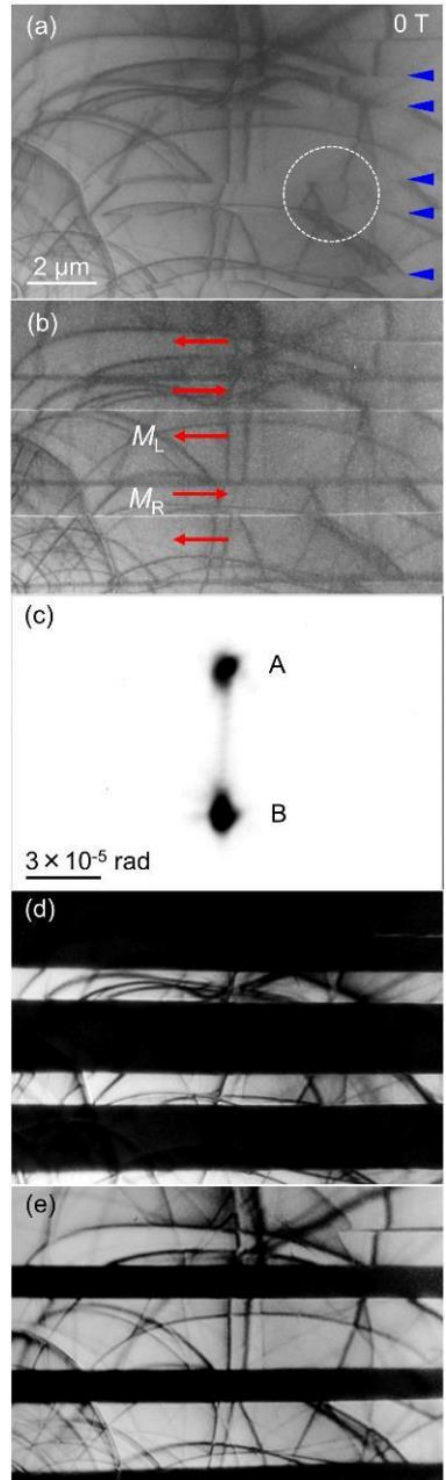

FIG. 3 (a) In-focus image of LSMO without an external magnetic field. Crystallographic twin boundaries are indicated by arrow-heads. (b) Under-focused Fresnel image. Arrows indicate the direction of magnetization in each magnetic domain. (c) SmAED pattern obtained from the region in the dotted circle of (a). (d) and (e) show Foucault images obtained by selecting spots A and B, respectively.

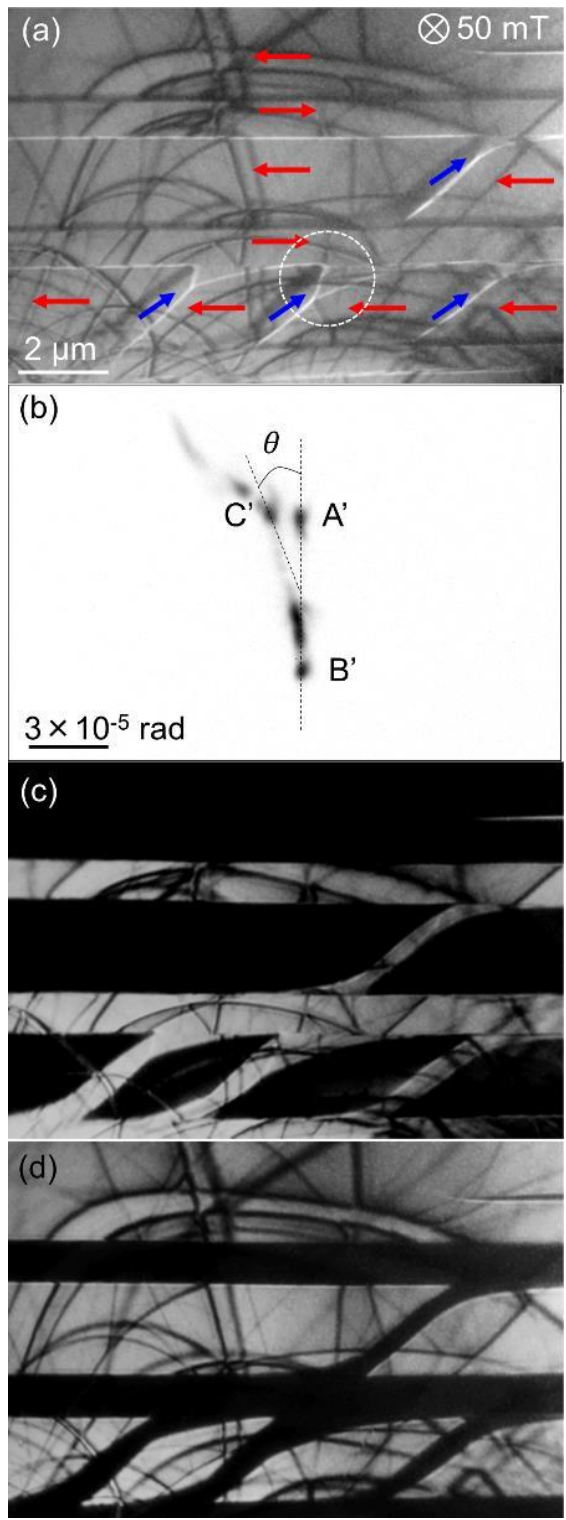

FIG. 4 (a) Under-focused Fresnel image of LSMO at 50 mT. (b) SmAED pattern obtained from the dotted white circular region. C' indicates a magnetically deflected spot due to the magnetic domains generated by applying an external magnetic field. (c) and (d) show Foucault images obtained by selecting two (A' and C') and one (B') of the split spots, respectively.

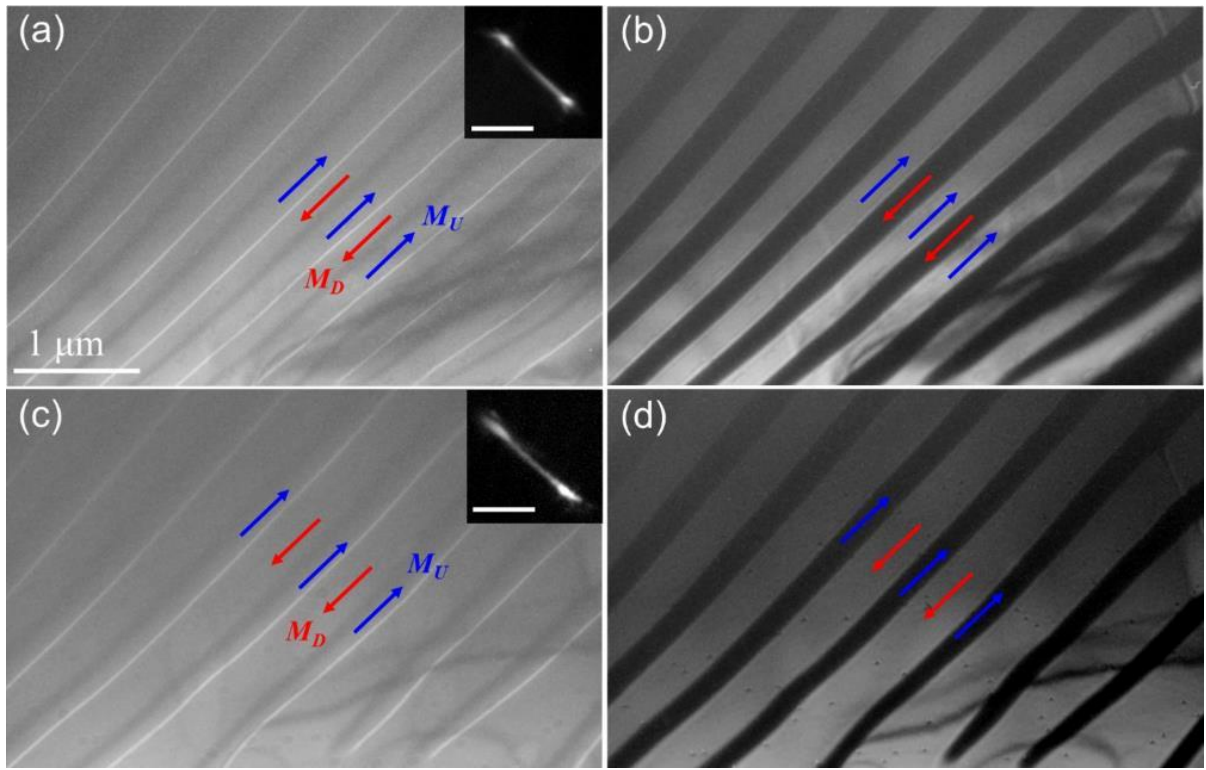

FIG. 5 (a) Fresnel image (under-focused) and (b) Foucault image (generated by selecting the left spot) of the ferromagnetic phase $Nd_{0.5}Sr_{0.5}MnO_3$ without an external magnetic field. (c) Fresnel image (under-focused) and (d) Foucault image at 170 mT. SmAED patterns are shown in the insets of (a) and (c). Scale bars in the insets are $5 \times 10^{-5}$ rad. The images are taken at 200 K.